\documentclass[aps,reprint,twocolumn,superscriptaddress,citeautoscript,showpacs,amsmath,amssymb,floatfix,prb]{revtex4-1}
\usepackage{graphicx}
\usepackage{amsmath}
\usepackage{amssymb}
\usepackage{mathtools}
\usepackage{textcomp,gensymb} 
\usepackage{bm} 
\usepackage[usenames,dvipsnames]{xcolor}
\usepackage{float}
\usepackage{setspace}
\usepackage{dcolumn}
\usepackage{array}

\usepackage{changepage}
\usepackage{tikz}
\usepackage{contour}
\usepackage[left]{lineno}
\usepackage{blindtext}
\usepackage{setspace}
\usepackage[none]{hyphenat}
\usepackage[colorlinks=true,citecolor=blue,linkcolor=blue,pdftex,hypertexnames=false,urlcolor=blue,linktocpage=true]{hyperref}
\definecolor{ashgray}{rgb}{0.7,0.75,0.71}
\definecolor{mspringgreen}{rgb}{0, 0.8, 0.1}
\definecolor{auburn}{rgb}{0.43, 0.21, 0.1}
\definecolor{ao(english)}{rgb}{0.0, 0.5, 0.0}
\definecolor{afw}{rgb}{0.95, 0.95, 0.96}
\definecolor{magnolia}{rgb}{0.97, 0.96, 1.0}
\definecolor{wsmk}{rgb}{0.96, 0.96, 0.96}

\newcommand{\gpt}{GaPt$_{5}$P}
\newcommand{\gmpt}{Ga$_{1-\mathrm{n}}$Mn$_n$Pt$_{5}$P}
\newcommand{\alpt}{AlPt$_{5}$P}
\newcommand{\alinpt}{(Al/In)Pt$_{5}$P}
\newcommand{\inpt}{InPt$_{5}$P}

\newcommand{\galpt}{Ga$_{1-x}$Al$_x$Pt$_{5}$P}
\newcommand{\ginpt}{Ga$_{1-x}$In$_x$Pt$_{5}$P}
\newcommand{\tmptpdpn}{TM(Pt/Pd)$_{5}$Pn}
\newcommand{\tmptpdp}{TM(Pt/Pd)$_{5}$P}

\newcommand{\xtpn}{XT$_{5}$Pn}

\newcommand{\mnpt}{MnPt$_{5}$P}
\newcommand{\xpt}{XPt$_{5}$P}

\newcolumntype{M}[1]{>{\centering\arraybackslash}m{#1}}

\usepackage{dcolumn}
\usepackage{multirow}

\usepackage{bbold}

\newcolumntype{d}[1]{D{.}{.}{#1}}

\begin{document}
\title{The first-order structural phase transition at low-temperature in \gpt\ and its rapid enhancement with pressure.}
\author{A. Sapkota}
\author{T. J. Slade}
\author{S. Huyan}
\author{N. K. Nepal}
\author{J. M. Wilde}
\author{N. Furukawa}
\affiliation{Ames National Laboratory, U.S. DOE, Iowa State University, Ames, Iowa 50011, USA}
\affiliation{Department of Physics and Astronomy, Iowa State University, Ames, Iowa 50011, USA}
\author{S. H. Laupidus}
\affiliation{Joint Center for Energy Storage Research and X-ray Science Division, Advanced Photon Source, Argonne National Laboratory, Lemont, Illinois 60439, USA}
\author{L.-L. Wang}
\affiliation{Ames National Laboratory, U.S. DOE, Iowa State University, Ames, Iowa 50011, USA}
\author{S. L. Bud$^{\prime}$ko}
\author{P. C. Canfield}
\affiliation{Ames National Laboratory, U.S. DOE, Iowa State University, Ames, Iowa 50011, USA}
\affiliation{Department of Physics and Astronomy, Iowa State University, Ames, Iowa 50011, USA}
\date{\today}

\begin{abstract}
Single crystals of \xpt\ (X = Al, Ga, and In), belonging to 1-5-1 family of compounds, were grown from a Pt-P solution at high temperatures, and measurements of the ambient pressure, temperature-dependent magnetization, resistivity, and X-ray diffraction were made. Additionally, the ambient-pressure Hall resistivity and temperature-dependent resistance under pressure were measured on \gpt. All three compounds have tetragonal $P4/mmm$ crystal structure at room-temperature with metallic transport and weak diamagnetism over the $2-300$~K temperature range. Surprisingly, at ambient pressure, both the transport and magnetization measurements on \gpt\ show a step-like feature in $70-90$~K region suggesting a possible structural phase transition. Neither \alpt\ nor \inpt\ have any signatures of a phase transition in their temperature-dependent electrical resistance and magnetization data. Both the hysteretic nature and sharpness of the features in the \gpt\ data suggest that the transition is the first order. Single-crystal X-ray diffraction measurements provided further details of the structural transition with a possibility of a crystal symmetry different than $P4/mmm$ below the transition temperature. The transition is characterized by anisotropic changes in the lattice parameters and a volume collapse with respect to the high-temperature tetragonal crystal structure. Furthermore, satellite peaks are observed at two distinct and non-equivalent wave-vectors (0, 0, 0.5) and (0.5, 0.5, 0.5), and density functional theory (DFT) calculations present phonon softening, especially at (0.5, 0.5, 0.5), as a possible driving mechanism. Additionally, we find that the structural transition temperature increases rapidly with increasing pressure, reaching room temperature by $\sim 2.2$~GPa, highlighting the high degree of pressure sensitivity of \gpt\ and fragile nature of its room-temperature structure. Even though the volume collapse and extreme pressure sensitivity suggest chemical pressure should drive a similar structural change in \alpt, where both unit cell dimensions and volume are smaller, its structure is found to be same as of the room-temperature \gpt. Overall, \gpt\ stands out as a sole member of the 1-5-1 family of compounds for which a temperature-driven structural changes has been observed. 

\end{abstract}

\maketitle
\section{Introduction\label{Int}}
Recent discoveries of diverse and novel magnetic properties in \tmptpdp\ (TM = 3d transition metals)\cite{Gui_2020,Gui_2021,Dissanayaka_2022,Tyler_2022,Tyler_2023_CrPtP,Tyler_2023_MnPdPt}, that can be easily tuned with chemical doping, have renewed interest in the 1-5-1 family of compounds. The 1-5-1 family of compounds, \xtpn\ (X =  transition metals, main group elements; T = Pd, Pt; Pn = P, As), were first reported in 1970 \cite{Boragay_1970} and the initial powder X-ray diffration (XRD) measurements on some members determined that they crystallize in a tetragonal structure with the space group $P4/mmm$. The tetragonal structure is characterized by the Pt and/or Pd atoms at $4i$ and $1a$, X atoms at $1c$ and Pn at $1b$ Wyckoff positions and is a layered structure with XPt$_{12}$ polyhedral layers separated along the $c-$axis by a Pn layer. 

Despite having the same tetragonal $P4/mmm$ crystal structure, the TM 1-5-1 compounds exhibit a diverse magnetism, from ferromagnetism (FM) to antiferromagnetism (AFM) and with local to itinerant characters\cite{Gui_2020,Gui_2021,Dissanayaka_2022,Tyler_2022,Tyler_2023_CrPtP,Tyler_2023_MnPdPt,Gui_FePt5P}. In particular, MnPt$_5$As\cite{Gui_2020} ($T_\mathrm{C} \approx 280$~K) and Cr$_{1+x}$Pt$_{5-x}$P\cite{Tyler_2023_CrPtP} ($T_\mathrm{C} \approx 464$~K)  exhibit ferromagentic ordering, whereas MnPt$_5$P\cite{Gui_2021} ($T_\mathrm{N} \approx 190$~K) and FePt$_5$P\cite{Gui_FePt5P} (multiple transitions between 70 and 90 K)  order antiferromagnetically. Among these, Cr$_{1+x}$Pt$_{5-x}$P and FePt$_5$P are itinerant, whereas MnPt$_5$P is local moment like. Also, isovalent doping of MnPt$_5$P with Pd, as detailed in Ref.~\citenum{Tyler_2023_MnPdPt}, results in a magnetic phase diagram characterized by multiple magnetic states and exemplifies chemical substitution as an effective tuning parameter of their physical properties.

Furthermore, the presence of a large amount of Pt or Pd can provide enhanced density of states at the Fermi surface and large spin-obit coupling, indicating a possibility of large, non-rare-earth based magnetocrystalline anisotropy. Overall, the \xtpn\ family of compounds offers a rich space to study the interplay between ferromagnetsim, antiferromagnetism, potential topological properties\cite{Tyler_2022} and possibly other structural and electronic phases\cite{Fujioka_2016}. To better understand the diverse properties exhibited by the $3d$ transition metal containing \xpt\ compounds, it is useful to provide a comparison with analogs in which X is not moment bearing.

 In this regard, 3A element (Al, Ga, In, Tl) \xpt\ members are ideal candidates as they do not have any moment bearing atoms and can be regarded as non-magnetic analogs of the \tmptpdpn\ compounds. Other than the first report of their structure in Ref.~\citenum{Boragay_1970}, only a few studies of some members of 3A element \xpt\ exist and the studies are mostly limited to the room temperature. For instance, powder XRD data on InPt$_5$P in Ref.~\citenum{zakharova_2018} confirmed its room-temperature crystal structure to be tetragonal with space group $P4/mmm$. Density Functional Theory (DFT) calculations predicted it to be metallic and diamagnetic, which was also confirmed by the magnetic measurements in the temperature range of $4-300$~K.

Here, we present temperature-dependence studies investigating the physical properties of three \xpt\ (X = Al, Ga, and In) compounds with the main focus on \gpt, due to an observed anomaly. Our powder X-ray diffraction results demonstrate that, similar to the earlier reports on \xtpn\ members, the room-temperature crystal structure of all three compounds is tetragonal with $P4/mmm$ crystal symmetry. Furthermore, the temperature-dependent electrical resistance and magnetic susceptibility of single crystalline \alpt\ and \inpt\ show metallic resistance versus temperature curves and weakly diamagnetic behavior (suggesting rather broad bands, or low density of states at the Fermi surface). No features suggesting any sort of phase transition exist in these data over the temperature range of $300$ to $2$~K. On the other hand, in addition to metallicity and diamagnetism, \gpt\ has a sharp and clear anomaly in both the electrical resistance and magnetic susceptibility in the $70-90$~K temperature range (depending upon the specific measurement as well as the direction of temperature change), indicating a possible structural change. In order to better understand the anomaly, additional measurements, such as single crystal X-ray diffraction, Hall resistivity, and high-pressure resistance measurements, were carried out on \gpt\ single crystals. 

Our single crystal X-ray diffraction measurements on \gpt\ confirm the first-order structural transition with anisotropic changes in lattice parameters, and a volume collapse with respect to the high-temperature tetragonal structure. The observed satellite peaks suggest a crystal symmetry with an enlarged unit cell, where $a^{\prime} = b^{\prime} = \sqrt{2}a$ and $c^{\prime} = 2c$, agreeing with the predictions from our DFT calculations. Furthermore, hysteresis, associated with the first-order structural transition, is apparent in all the measurements. The low temperature structure of \gpt\ is categorized as a double-Q type structure with respect to $P4/mmm$, as evidenced by the presence of satellite peaks with two distinct wave-vectors. DFT calculations present phonon softening as a possible mechanism for the transition. 

In addition, we find that the structural phase transition in \gpt\ is remarkably pressure sensitive, as the resistive feature associated with it shifts towards higher temperature with increasing pressure, bringing it to $\sim 300$~K by $2.2$~GPa. Given that the structural phase transition temperature crosses the pressure medium solidification line, \gpt\ offers a case study of the effects of pressure medium hydrostaticity on a structural phase transition.

\section{Experimental Details and computational methods\label{ED}}
\subsection{Crystal Growth\label{CG}}
Single crystals of \xpt\ (X = Al, Ga, In) were grown from a ternary X-Pt-P solution analogous to that used to produce other MPt$_5$P phases (M = Cr, Mn, Fe)\cite{Tyler_2022,Tyler_2023_CrPtP}. The starting materials were elemental Pt powder (Engelhard, 99+ $\%$ purity), chips of red P (Alpha Aesar, 99.99 $\%$) and  Ga or Al or In (Alpha Aesar, 99.999 $\%$). The elements were weighed according to a nominal molar ratio of X$_9$Pt$_{71}$P$_{20}$ and contained in the bottom side of a 3-piece alumina Canfield crucible set (CCS, sold by LSP Ceramics)\cite{CCS,Canfield_crucible}. The CCS was sealed in a fused silica ampule and was held in place with a small amount of silica wool, which serves as cushioning during the decanting step\cite{Canfield_2020}. The ampules were evacuated three times and back-filled with $\sim 1/6$ atm Ar gas prior to sealing. Using a box furnace, the ampules were first warmed to 250 \degree C over 6 h, then to 1180 \degree C in an additional 8 h. After holding at 1180 \degree C for 6 h, the furnace was cooled to 830 \degree C in 72 h, and upon reaching the final temperature, the remaining liquid phase was decanted by inverting the ampules into a centrifuge with specially made metal rotor and cups\cite{Canfield_2020}. After cooling to room temperature, the ampule was cracked open to reveal clusters of thin, metallic, plate-like crystals. Some of the representative single crystals of \gpt\ grown using this method are shown in Fig.~\ref{RCS}(c), below, over a mm grid. Single crystals of \alpt\ and \inpt\ were similar, although the \alpt\ crystals were not as well faceted.

\subsection{Powder X-ray Diffraction\label{PXRD}}
The room temperature crystal structures of the three \xpt\ compounds were confirmed from the refinement of the powder X-ray diffraction data measured at 11-BM beamline of the Advanced Photon Source at Argonne National Laboratory. Select single crystals of \alpt, \gpt, and \inpt\ were ground with a mortar and pestle to a fine powder and passed through using a $40~\mu m$ mesh sieve. Each of the powders were loaded into a 11-BM Kapton capillary tube and the diffraction patterns were collected using monochromatic X-ray radiation of wavelength of $\lambda = 0.4589$~\AA\ from -6 to 28\degree\ 2$\theta$ with a step size of 0.001\degree\ and with counting time of 0.1 s/step. Rietveld refinement using GSAS-II software\cite{Toby_2013_GSAS} was performed to confirm the reported crystal structure.

\subsection{High-resolution single crystal X-ray diffraction}
Single crystal X-ray diffraction measurements were performed on an in-house four-circle diffractometer (Huber) using Cu $K_{\alpha 1}$ radiation from a rotating anode X-ray source, using a germanium (1 1 1) monochromator. A He closed-cycle refrigerator was used for temperature dependent measurements between 10 and 120 K. Three beryllium domes were used as a vacuum shroud, heat shield, and the last innermost dome containing the sample. The innermost dome was filled with a small amount of He gas to improve thermal contact to the sample surface. Measurements were carried out on single crystal of \gpt\ with mass $\sim 0.025$ g attached to a flat copper sample holder that is attached to the cold finger. Additionally, room-temperature measurements were carried on \alpt\ and \inpt\ single crystals with masses of $\sim0.004$ and $\sim0.002$ g, respectively.

\subsection{Magnetization and Transport}
Magnetization measurements were performed in a Quantum Design Magnetic Property Measurement System (QD MPMS-classic) SQUID magnetometer operating in the DC measurement mode. The magnetic measurements were conducted with the field oriented parallel and perpendicular to the c-axis, where c is perpendicular to the large surface of the plate-like crystals \big[see Fig.~\ref{RCS}(c) below\big]. For the H $\parallel$ c measurements, the sample was glued to a plastic disc, and background measurements using the empty disc were first measured and the values subtracted. For the H $\perp$ c measurements, the sample was held in place between two straws.

Resistance measurements were done with a Quantum Design Physical Property Measurement System (QD PPMS). The samples were prepared by cutting the crystals into rectangular bars, and the contacts were made by spot welding $25~\mu m$ thick annealed Pt wire onto the samples in standard, linear four point geometry so that the current was flowing within the ab-plane. To ensure good mechanical strength, a small amount of silver epoxy was painted over the spot-welded contacts, and the contact resistances were normally $\sim~1~\ohm$. 

Hall resistivity was measured using a four-probe configuration with the applied field held perpendicular to the current direction. To account for small misalignment of the Hall voltage contacts, the odd component of the signal $\rho_\mathrm{H} = \dfrac {\rho(+H) - \rho(-H)}{2}$, was taken as Hall resistivity. Here, $\rho(+H)$ and $\rho(-H)$ are the measured Hall resisitivity values for positive and negative applied magnetic field, respectively. Electrical transport measurements were performed using an ACT option of the QD PPMS.

\subsection{Transport under pressure}

The temperature dependent electrical transport measurements down to 1.8 K and under pressures up to 2.2 GPa were carried out into our a lab-made piston-cylinder pressure cell\cite{Budko_1984} that fits QD PPMS. A linear, 4-terminal configuration was used for the arrangement of wires for contacts, with the applied current direction along ab plane.  High-purity lead (Pb) was used for the determination of pressure at low temperatures\cite{Eiling_1981}.  The mixed fluid with 4:6 of light mineral oil and n-pentane was used as the pressure transmitting medium, which solidifies at room temperature in the pressure range of 3–4 GPa and has a well determined T(P) solidification line\cite{Torikachvili_2015}. 

The \gpt\ single crystal used for these measurements under pressure is different than the ones used for the ambient-pressure transport measurements discussed in the previous section.

\subsection{Computational method}

Density functional theory \cite{Hohenberg_1964,Kohn_1965} (DFT) calculations were performed with PBEsol exchange-correlation functional including spin-orbit coupling (SOC) using a plane-wave basis set and projector augmented wave method\cite{Blochl_1994}, as implemented in the Vienna Ab-initio Simulation Package\cite{Kresse_1996,Kresse_1996b} (VASP). We used a kinetic energy cutoff of $319$~eV, a $\Gamma$-centered Monkhorst-Pack\cite{Monkhorst_1976} (10×10×6) k-point mesh, and a Gaussian smearing of $0.05$~eV. The ionic positions and unit cell vectors were fully relaxed with the remaining absolute force on each atom being less than $2 \times 10^{-3}$ eV/Å. The DFT-calculated lattice parameters with PBEsol+SOC are within $1\%$ of the experimental data. Phonon band dispersion of the \gpt\ were calculated with finite difference method using PHONOPY\cite{Togo_2015}.

\section{Results and Discussion\label{RS}}
\subsection{Transport and magnetic properties of \xpt\ (X = Al, Ga, In) \label{TM3A}}

\begin{figure}[]
	\centering
	\includegraphics[scale=0.7]{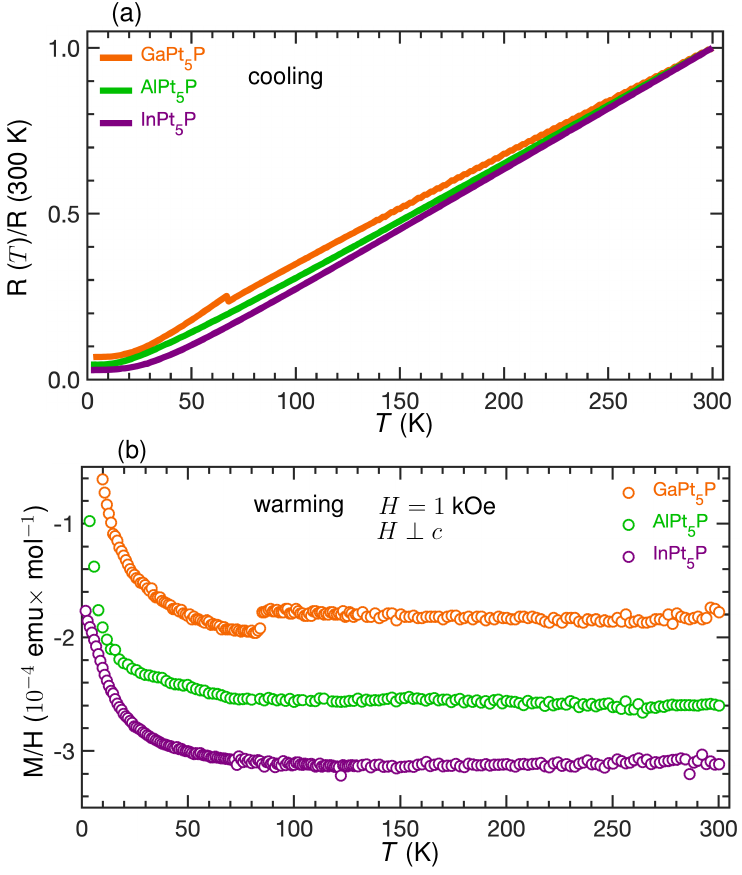}
	\caption{Temperature dependence of (a) resistance $R$ normalized to its 300 K value and (b) susceptibility ($\chi = \dfrac {M} {H}$) for \alpt, \gpt, and \inpt. The measurements cover the temperature range of $\sim 2$ to 300 K and include cooling data for the resistance and warming for magnetization.} 
	\label{MRT_all}
\end{figure}

Figure~\ref{MRT_all} shows the temperature dependence of the normalized resistance (while cooling) and magnetic susceptibilities (while warming) for \alpt, \gpt, and \inpt. All three compounds exhibit metallic temperature dependence of the resistance and essentially temperature independent magnetic susceptibilities (except for small, low-temperature Curie tails, to be discussed below), consistent with a low density of states at the Fermi surface. DFT calculations of the electronic structures of \gpt, shown in Fig.~\ref{DFTf} of Section~\ref{DFTC}, below, indicate small but non-zero density of states (DOS) at the Fermi level, and similar results were reported for \inpt\ in Ref.~\citenum{zakharova_2018}. The main contribution at the Fermi level is due to the \textit{d}-orbitals of the transition metal Pt. 

The residual resistivity ratios $RRR = R(300 K) / R(2.0 K)$ are 20, 15, and 34 for \alpt, \gpt, and \inpt, respectively, and are consistent with well-ordered compounds. Overall, the magnetic susceptibilities display features typical of non-moment bearing compounds with a small amount of paramagnetic (PM) impurity. To quantify the level of the PM impurities, we fit the low-temperature part ($< 50$~K) of the data using Curie-Weiss law,
 \begin{equation}\label{CW}
 \chi = \dfrac{C}{T-T_{c}} + \chi_0 
 \end{equation}
 where $C$ is Curie constant; $C = 0.002$ emu K/mol was obtained from our fit of the \gpt\ data in Fig.~\ref{MRT_all}(b). As discussed in Ref.~\citenum{CurieWeiss_2022}, the value of Curie constant $C$ can be used to calculate the effective magentic moment per formula unit (f.u.) using the relationship,
\begin{equation}\label{mueff}
\mu_\mathrm{eff} = \sqrt{\dfrac{8C}{n}} \mu_\mathrm{B} [cgs]
\end{equation}
where $n$ is number of magnetic atoms per formula unit. 

Next to gauge the amount of the impurity, we assumed that the magnetic impurity present in our \gpt\ sample is Mn$^{2+}$ (just an assumption not a known impurity) and exist as \gmpt, given that \mnpt\ is a known member of 1-5-1 family. The spin only effective moment of Mn with S = 5/2 and $g = 2$ can be calculated using,
\begin{equation}\label{mucalc}
\mu^{\mathrm{Mn}}_\mathrm{cal} = g\sqrt{S(S+1)} \mu_\mathrm{B} = 5.9 \mu_\mathrm{B}
\end{equation}

Replacing this value in Eq.~\ref{mueff} as $\mu_\mathrm{eff}$, we get $5.9 \mu_\mathrm{B}=\sqrt{\dfrac{8C}{n}} \mu_\mathrm{B}$ and on solving this, we get $n = 0.0005$, i.e. very small amount of Mn impurity. Even for the smaller moment impurity with $S = 1/2$ and $g = 2$, the impurity concentration is still very small, $n = 0.005$.
Overall, the implication of our calculation is that the amount of the paramagnetic impurities (PMI) present in our sample Ga$_{1-\mathrm{n}}$(PMI)$_n$Pt$_5$P were infinitesimal. This low level of impurities, or disorder, is consistent with the relatively large RRR values we found above.

Also, the Curie-Weiss fits to the low-temperature part ($< 50$~K) of the $\dfrac {M}{H}$ data of \alpt\ and \inpt, shown in Fig.~\ref{MRT_all}(b), produce C = 0.0007 and 0.0028 emu K/mol, respectively. The values are comparable to that of \gpt\ indicating infinitesimal amount of PMI in these samples as well.

Besides metallicity and diamagnetic behavior, \gpt\ has a conspicuous, anomalous feature that is clearly visible as a sharp jump in both resistance and susceptibility data. The jump appears near $70$~K in the resistance (cooling) data and near $84$~K in the susceptibility (warming) data, suggesting a first order phase transition with rather large hysteresis. Given that no such transitions exist for both \alinpt\ and have not been reported for any of the known \xpt\ materials, we focus on exploring the feature in \gpt\ for the rest of the paper.

\subsection{Transport and magnetic properties of \gpt: A first-order structural transition \label{TMGPT}}

\begin{figure}[]
	\centering
	\includegraphics[scale=0.75]{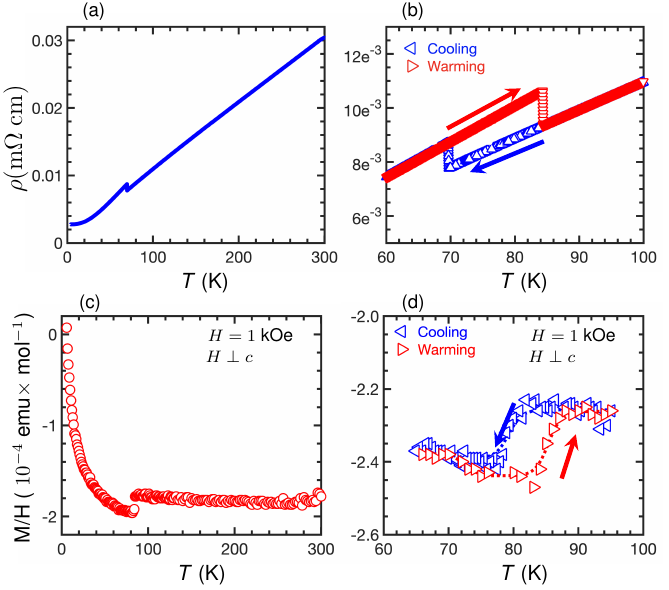}
	\caption{Temperature dependence of resistivity and susceptibility ($\chi = \dfrac {M} {H}$) of \gpt. (a) and (c) show the measurements in the large temperature range up to 300 K and contains either cooling or warming data, respectively. (b) and (d) show the results for both warming and cooling to highlight the hysteresis, and contains data only around the anomalous jump or transition.}
	\label{MRT}
\end{figure}

Figure~\ref{MRT} shows the temperature dependence of the resistivity and magnetic susceptibility ($\chi = \dfrac {M} {H}$) for \gpt\, only. As discussed above, besides metallicity and diamagnetism, sharp and discontinuous changes are apparent in both the measurements, and the onset for warming and cooling occurs at different temperature indicating the hysteresis. Onset here corresponds to the starting point of the discontinuity and the values are different for warming and cooling measurements.

\begin{figure}[]
	\centering
	\includegraphics[scale=0.7]{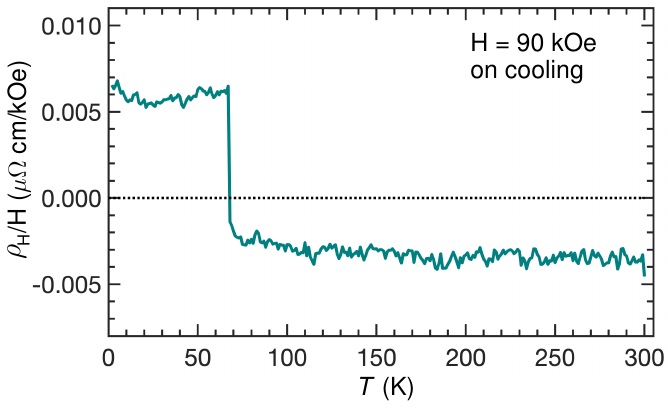}
	\caption{Temperature dependence of the Hall coefficient ($\rho_\mathrm{H}/H$) of \gpt.}
	\label{HC}
\end{figure}

In case of the resistivity, as shown in Fig.~\ref{MRT}(b), the sharp, discontinuous increase appears upon cooling near $70$~K, and the higher-temperature values are recovered on warming near 84 K. The hysteresis between warming and cooling profiles implies a first-order transition. Alteration to the electronic structure induced by a structural transition is the likely cause of the resistive anomaly. The temperature-dependent Hall data, shown in Fig.~\ref{HC}, indeed supports this notion of a change in the electronic structure. Below $\approx 70$~K, the Hall coefficient also changes abruptly from negative to positive, indicating the dominant charge carrier changes from the electron to the hole type. 


Like the resistivity curves, the hysteresis observed in the $\dfrac {M}{H}$ data, in Fig.~\ref{MRT}(d), supports the idea that the transition is first-order. Also, we note that the width of the hysteresis between warming/cooling sweeps is slightly wider in the resistance measurement, which most likely reflects the faster warming-cooling (ramp) rate used for these measurements than for the magnetization. Moreover, the hysteresis width and transition temperatures exhibit history dependence in \gpt. Variation in the hysteresis width and transition temperatures of \gpt\ with both the ramp rate and in different cycles of the measurements is shown in Fig.~\ref{MTDC}, below, in the Appendix and discussed briefly in Section~\ref{HDHTT}. 

The change in magnetization at the transition is consistent with the change in the Hall coefficient. The larger, diamagnetic signal, just below the transition [Figs.~\ref{MRT} (c) and (d)], implies a decrease in the Pauli paramagnetism which itself would be associated with a decrease in DOS ($E_\mathrm{F}$).  Such a change is also implied by the Hall data which has a smaller magnitude for temperatures just above the transition temperature and a larger magnitude just below. As such, the Hall data also imply a smaller DOS ($E_\mathrm{F}$) for temperature below the transition temperature.

\subsection{Powder X-ray diffraction: Room-temperature crystal structure\label{RTCS}}

\begin{figure*}[]
	\centering
	\includegraphics[scale=0.85]{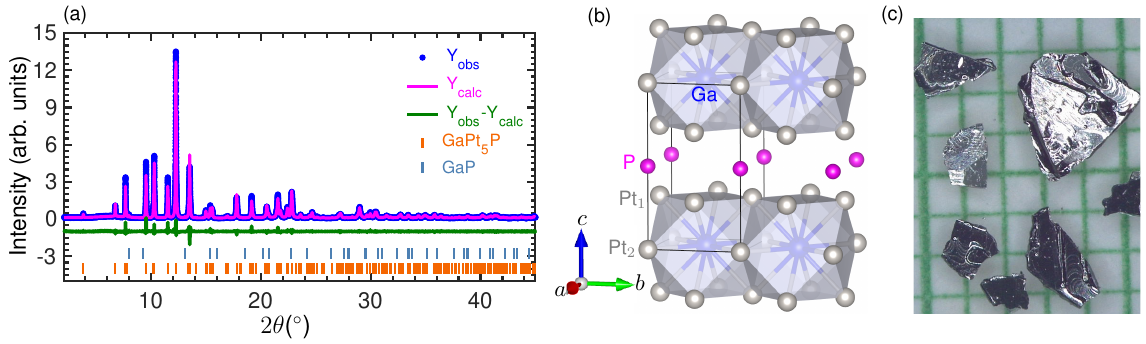}
	\caption{(a) Reitveld refinement of the room-temperature synchrotron powder X-ray diffraction data of \gpt\ with $R_{wp} = 9.87\%$, $R_{p} = 7.97\%$ and GOF $= 1.56$. Blue dots and magenta solid lines correspond to the data and calculated pattern, respectively. The green line and vertical orange markers are the difference between data and calculation and allowed Bragg reflections, respectively. (b) Crystal structure of \gpt. (c) Pictures of solution grown \gpt\ single crystals on a mm grid. The crystallographic $c-$axis is perpendicular to the plate-like surfaces.}
	\label{RCS}
\end{figure*}
Before studying \gpt\ structure upon cooling, it is important to more carefully examine its structure at room temperature and compare it to \alpt\ and \inpt, especially considering the compelling evidence indicating a possible structural phase transition in \gpt\ below room temperature. 

Hence, to confirm the room-temperature crystal structure, our synchrotron powder X-ray diffraction data measured on \gpt\ were refined using Rietveld analysis and GSAS-II software. In agreement with the previous reports\cite{Boragay_1970}, the room temperature crystal structure of \gpt\ is tetragonal with space group $P4/mmm$. Room-temperature powder X-ray diffraction data and the final refinement plots are shown in Fig.~\ref{RCS}(a). The crystal structure obtained from the refinement is shown in Fig.~\ref{RCS}(b) and is isostructural to other \xpt\ compounds. It is a layered structure, comprising of a cuboctahedron of Ga-centered GaPt$_{12}$ layer separated along the \textit{c}-axis by a P layer.

\begin{table}[]
	\caption{Unit cell parameters at room temperature obtained from the Rietveld refinement.\label{UCP}}
	\begin{ruledtabular} 
		\begin{tabular}{ccc}
            Formula & & \gpt \\
            space group & & $P4/mmm$ \\
			$a$ (\AA) & &3.90038(2)\\
			$c$ (\AA) & &6.85564(3)\\
            $\alpha$ & &90\\
            $\beta$  & &90\\
            $\gamma$ & &90\\
            volume (\AA$^{3}$) & &104.2948(8)\\
		\end{tabular}
	\end{ruledtabular}
\end{table}

In addition to the main phase, a secondary phase of GaP exists in a small amount ($ < 0.5~\mathrm{wt.}~\%$). Statistical parameters from the refinement including both the phases in the model are $R_{\mathrm{wp}} = 9.87\%$ and GOF$ = 1.56$ indicating a reasonable refinement. The unit cell parameters, atomic coordinates and isotropic displacement parameters obtained from the refinement are summarized in Table~\ref{UCP} and \ref{ACEID}. Refined parameters are similar to the ones reported earlier for \gpt\ in Ref.~\citenum{Boragay_1970}.

 \begin{table}[]
	\caption{Refined atomic coordinates and the equivalent isotropic displacement parameters of \gpt\ at room temperature.\label{ACEID}}
	\begin{ruledtabular} 
		\begin{tabular}{ccccccc}
           Atoms& Wyckoff& $x$& $y$& $z$& Occ.& $U_\mathrm{eq}$\\
           \hline
           Pt1 &$4i$& 0 & 0.5 &0.28978(3)& 1 & 0.0043(1)\\
           Ga &$1c$& 0.5 & 0.5 &0& 1 & 0.0080(5)\\
           P &$1b$& 0 & 0 &0.5& 1 & 0.0092(11)\\
           Pt2 &$1a$& 0 & 0 & 0 & 1 & 0.0045(2)\\
		\end{tabular}
	\end{ruledtabular}
\end{table}

\begin{figure}[]
	\centering
	\hspace*{-4mm}
	\includegraphics[scale=0.72]{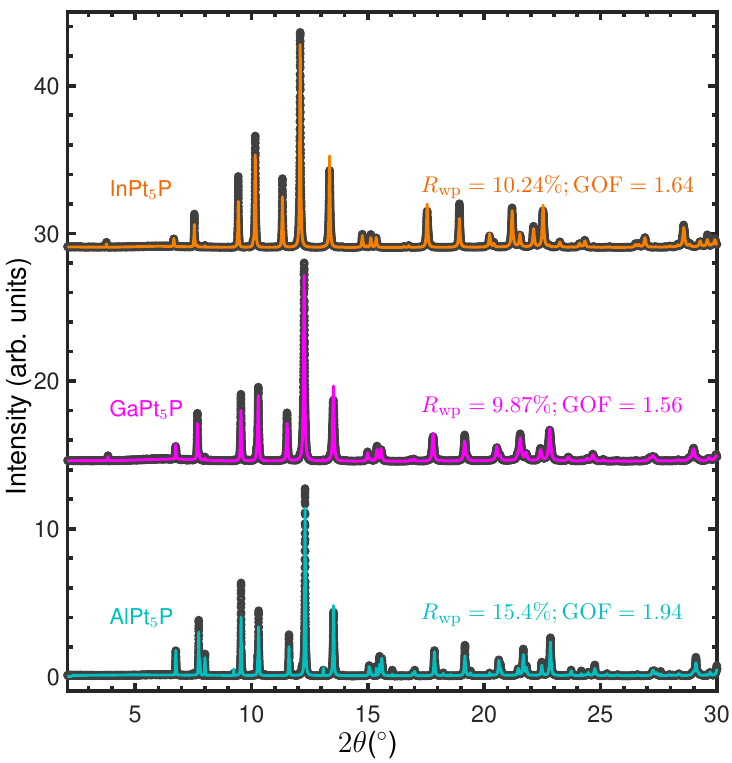}
	\caption{Rietveld refinement of the room-temperature synchrotron powder X-ray diffraction data of \xpt\ X = Al, Ga, and In. Black dots and colored solid lines correspond to the data and calculated pattern, respectively. Corresponding $R_{wp}$ and GOF are given alongside with the figure with same color as the calculated pattern. Small minority peaks corresponding to the binary compounds AlP, GaP and InP were present in the small amount in all 3 powder patterns, respectively, but in case of \alpt\, additional impurity peaks corresponding to other binary, such as PtP$_2$, were also present. Measurements on all three powder samples were carried out with identical conditions as discussed in Section~\ref{PXRD}.}
	\label{3pwd}
\end{figure}

Similar synchrotron powder X-ray measurements were also carried out on \alpt\ and \inpt\ for comparison. Figure~\ref{3pwd} shows the powder pattern for all 3 compounds along with their respective fits from the refinement with $P4/mmm$ crystal symmetry. R-factors and the goodness of fits, listed in the figure, obtained from all 3 refinements indicate a reasonable refinement, thereby confirming that the room-temperature crystal structure for all 3 (Al, Ga, In)Pt$_5$P compounds is indeed tetragonal with the space group $P4/mmm$. Refined lattice parameters for all 3 compounds are listed in Table~\ref{ACALL}. The lattice parameters and volume increase from Al to In, in accordance with the corresponding increase in atomic radii from Al to In.

\begin{table}[]
	\caption{Refined lattice parameters and volume of \alpt, \gpt\ and \inpt\ at room temperature.\label{ACALL}}
	\begin{ruledtabular} 
		\begin{tabular}{cccc}
             & $a$ (\AA) & $c$ (\AA)& Volume (\AA$^{3}$)\\
           \hline
           \alpt&3.89832(1) & 6.80627(3) & 103.434(1)\\
            \gpt & 3.90038(2) & 6.85564(3) & 104.295(1)\\
           \inpt&3.94700(1) & 6.97401(3) & 108.647(1)\\
		\end{tabular}
	\end{ruledtabular}
\end{table}

Impurities or secondary phases identified in \alpt, \gpt\ and \inpt\ samples from our powder diffraction measurements are binary AlP, GaP and InP, respectively. All three of them are diamagnetic\cite{Fan_2019_AlNM,Haynes_2019,Sahu_InP_DM,Dean_1977} as well as the PtP$_2$\cite{Baghdadi_1974_PtPDM} found in \alpt. Hence, the paramagnetic impurities suggested from the susceptibility plots, in Figs.~\ref{MRT_all}(b), and discussed above in Section~\ref{TM3A}, are yet to be identified. Absence of the peaks corresponding to the paramagnetic impurities are due to their infinitesimal amount in the sample.

\subsection{Single crystal X-ray diffraction of \gpt: Structural transition with a potential crystal symmetry change\label{SXRD}}

\begin{figure}[]
	\centering
	\hspace*{-5mm}
	\includegraphics[scale=0.75]{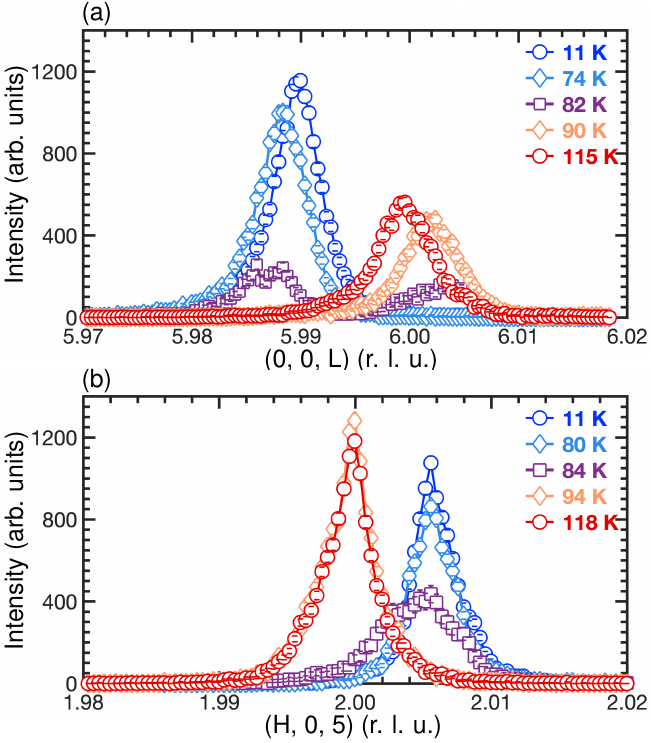}
	\caption{(a) and (b) Evolution of the (0, 0, 6) and (2, 0, 5) Bragg peaks of \gpt\ with temperature while warming.}
	\label{BP}
\end{figure}

To further determine the details of the structural transition in \gpt, we performed a high-resolution single crystal X-ray diffraction measurements between 10 to 120 K. Various points in the reciprocal space (H 0 L) with respect to tetragonal $P4/mmm$ crystal symmetry were measured, and to acquire the lattice parameters, (0, 0, 6) and (2, 0, 5) Bragg peaks were measured on warming and cooling. First, the (0 0 6) peak was measured in the ramp-mode with rate of 0.1 K/min and the measurements of (2, 0, 5) peaks after with the rate of 0.15 K/min. The slow ramp rates minimize the temperature lag between the sensor and sample. Figure~\ref{BP} shows the evolution of the (0, 0, 6) and (2, 0, 5) peaks while warming. Both peaks show a sudden shift in the peak positions upon warming through the mid-80 K's. At $82$~K, a splitting of the (0, 0, 6) peak is apparent, and at 84 K, a shoulder on the (2,0,5) peak is observed. These features indicate the coexistence of two phases and support the notion of the first-order structural transition.

\begin{figure}[]
	\centering
	\hspace*{-5mm}
	\includegraphics[scale=0.7]{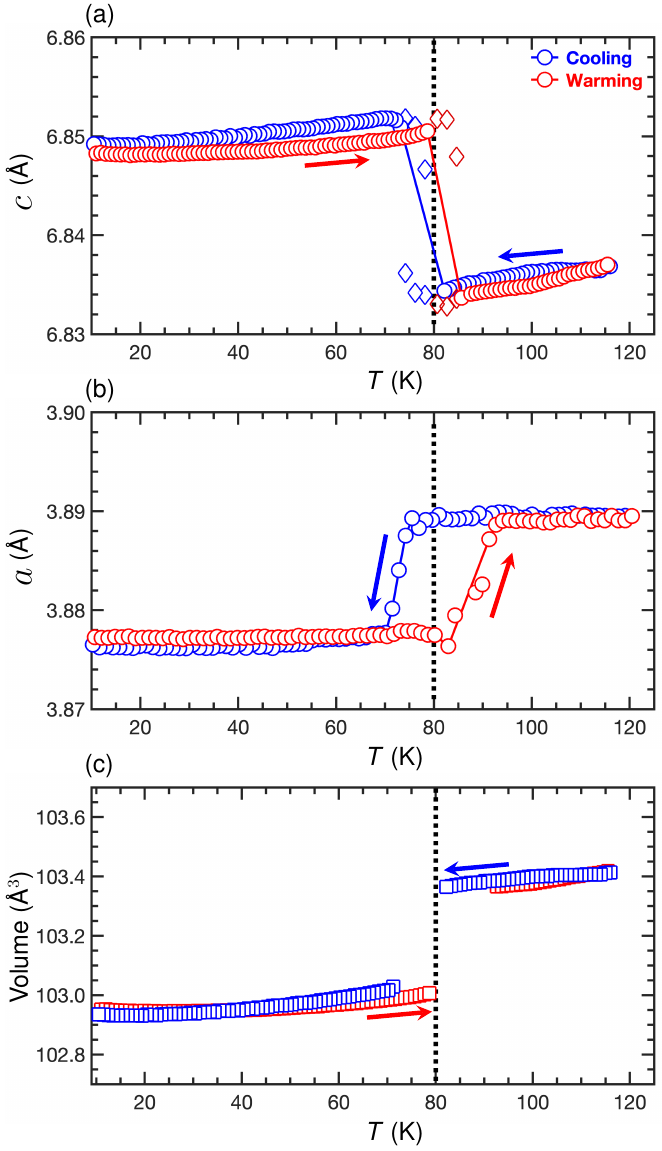}
	\caption{Temperature dependence of the lattice parameters and volume, defined with respect to high-temperature tetragonal structure, while cooling and warming. Blue and red data correspond to measurements while cooling and warming, respectively. Diamond markers in (a) correspond to the coexistence region, hence the figure has two values. Solid blue and red lines in (a) and (b) connect the data points, excluding the coexistence regions, and provide a qualitative visualization of the hysteresis. Additionally, the dotted vertical line at $\sim 80$~K is the onset temperature, at least for the warming measurements, and roughly highlights the transition temperature $T_\mathrm{s}$ in our measurements.}
	\label{acvT}
\end{figure}

Next, the lattice parameters of the tetragonal unit cell were obtained from the fits of the peaks in Fig.~\ref{BP} with a Pseudo-Voigt lineshape. Consistent with the first-order transition, lattice parameters in Fig.~\ref{acvT} show abrupt and discontinuous changes along with clear thermal hysteresis. However, the changes in both lattice parameters are anisotropic: the $c$ lattice parameter increases below the transition temperature whereas the $a$ lattice parameter decreases. For warming, the onset for the transition is $T_\mathrm{onset} \sim 80$~K for both $c$ and $a$ lattice parameters but differs in the case of cooling. The temperature evolution of the volume obtained by using these lattice parameters, except in the coexistence region, is shown in Fig.~\ref{acvT}(c) and upon cooling the unit cell volume decreases by $0.33\%$. Hysteresis is apparent in each of the three figures, but the hysteresis width differs for $c$ and $a$ lattice parameters, Figs.~\ref{acvT}(a) and (b), respectively. As shown in Fig.~\ref{MTDC}, below, in Section~\ref{HDHTT} of Appendix, the difference most likely arises from the combination of different ramp rates and the measurements being done during different cycles.

\begin{figure}[]
	\centering
	\includegraphics[scale=0.75]{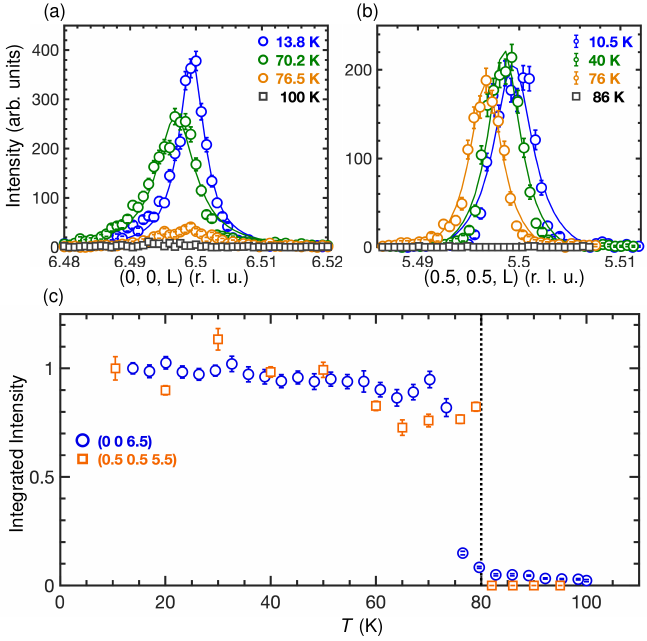}
	\caption{X-ray diffraction measurements of the satellite peaks at wave-vectors (a) $\tau = $ (0, 0, 0.5) and (b) $\tau = $ (0.5, 0.5, 0.5) with respect to the high-temperature $P4/mmm$ crystal symmetry. (a) [0, 0, L] scans of the (0, 0, 6.5) Bragg peak on cooling illustrating its evolution. (b) Similar plots for the (0.5, 0.5, 5.5) Bragg peaks measured while warming. (c) Normalized integrated intensity of the (0, 0, 6.5) and (0.5 0.5 5.5) satellite peaks as an order parameter showing the discontinuity and phase transition below $\approx 80$~K on warming (with a lower temperature discontinuity on cooling). The integrated intensities are normalized to the corresponding value at the base temperature and are obtained after fitting the plots in (a) and (b) with pseudovoigt lineshapes.}
	\label{SL}
\end{figure}

In addition to the changes observed in the lattice parameters and nuclear Bragg peaks, satellite peaks at the commensurate wave-vectors of $\tau = $ (0, 0, 0.5) and (0.5, 0.5, 0.5) were observed below the transition. Figures~\ref{SL}(a) and (b) show the temperature evolution of the satellite peaks at wave-vectors (0, 0, 6.5) and (0.5, 0.5, 5.5), respectively. Both figures clearly illustrate their appearance at lower temperature approximately below transition temperature $T_\mathrm{s}$ in Fig.~\ref{acvT}. To further highlight this, integrated intensity from the fits of the peaks in (a) and (b), is shown in Fig.~\ref{SL}(c). Similar to the lattice parameters, we found that the forbidden satellite peaks appear below $T_\mathrm{s} \approx 80 $~ K (on warming) and also have a discontinuity or sharp onset. Furthermore, these superlattice peaks are not only found at a single point in the reciprocal space but were also measured in multiple Brillouin zones, as shown in Fig.~\ref{SL_diffBZ} in Section~\ref{SPBZ} in Appendix, below.

Appearance of additional peaks at the forbidden positions indicate that the volume not only collapses below the transition, as shown in Fig.~\ref{acvT}(c), but there is a possibility of the structure transforming to a different crystal symmetry than the high-temperature $P4/mmm$. Furthermore, peaks at $\tau = $ (0, 0, 0.5) and (0.5, 0.5, 0.5) positions indicate the doubling of $c$ as well as the enhanced in-plane unit cell dimensions, with potentially $a^{\prime} = b^{\prime} = \sqrt{2}a$, similar to the DFT predictions in Section~\ref{DS}. Other possible scenarios, like $a^{\prime} = b^{\prime} = 2a$, can be ruled out because the observed feature in our (H, 0, L) Bragg peaks is inconsistent with the lower crystal symmetry expected for this case. For instsance, ISODISTORT\cite{Campbell_2006} analysis of the potential subgroups, considering two \textbf{k}-vectors (0, 0, 0.5) and (0.5, 0.5, 0.5) and their superposed irreducible representations (Irreps) suggest an orthorhombic crystal structure for $a^{\prime} = b^{\prime} = 2a$ and should lead to the splitting of (2, 0, 5) peak below transition, but no such splitting was observed in our data, as shown in Fig.~\ref{BP}. Additionally, in our measurements, we also checked for the splitting along [H, H, 0] direction by measuring (1, 1, 6) Bragg peaks and no such splitting were found. The results imply that the low-temperature crystal structure is potentially tetragonal because for other lower symmetry, splitting of the tetragonal Bragg peaks are expected due to formation of the structural domains. Finally, since the satellite (superlattice) peaks were observed at two distinct and non-equivalent wave-vector positions, we categorized the low-temperature structure as double-Q structure with respect to room-temperature tetragonal $P4/mmm$. 

Overall, our single crystal X-ray diffraction measurements confirm that \gpt\ undergoes a first-order structural transition from tetragonal $P4/mmm$ to a structure with potentially a different crystal symmetry, and with an enlarged unit cell of $a^{\prime} = b^{\prime} = \sqrt{2}a$ and $c^{\prime} = 2c$. Additionally, the transition is characterized by the anisotropic changes of lattice parameters and a volume collapse with respect to the high-temperature tetragonal structure.

\subsection{Transport under pressure in \gpt: Enhancement of $T_\mathrm{s}$ towards room-temperature\label{TP}}
The apparent change in the unit cell volume below the transition indicates that it should be possible to couple the transition to applied hydrostatic pressure. As a matter of fact, in a simplistic model, taking into account the volume collapse, hydrostatic pressure should enhance or promote the transition by driving $T_\mathrm{s}$ to the higher values. However, the anisotropic changes in the $c$ and $a$ lattice parameters allow for the possibility that this assumption might be oversimplified. Hence, to understand how the pressure affects the observed transition and to quantify its sensitivity, zero-field resistance measurements between $1.8-300$~K were carried out on \gpt\ single crystals at different pressures using a lab made piston-cylinder pressure cell. The measurements, including the ambient pressure measurement outside the pressure cell, were done on a single crystal different than the one used for the measurements in Figs.~\ref{MRT_all} and \ref{MRT}.

\begin{figure}[]
	\centering
	\hspace*{0mm}
	\includegraphics[scale=0.7]{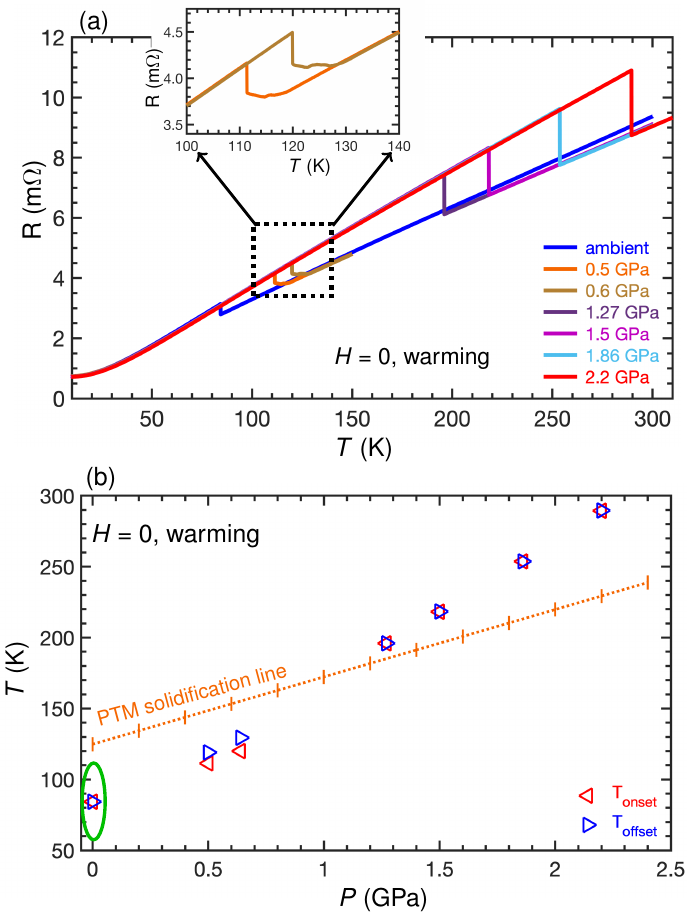}
	\caption{(a) Temperature dependence measurements of the resistance at different pressures illustrating the increase in the feature (upturn or jump) corresponding to the structural transition. The inset is a magnified plot, zoomed in to highlight the broadening of an otherwise sharp upturn. (b) Pressure-temperature (P-T) phase diagram of \gpt\ from $H = 0$ resistance measurements showing the onset and offset temperatures in (a). In addition, it contains the solidification line (dotted orange with vertical dashes) of the pressure-transmitting media (light oil and n-pentane) at different pressures taken from the Ref.~\citenum{Torikachvili_2015}. The green ellipse around the $P \sim 0$~GPa data is to emphasize that the ambient pressure measurements were done outside the pressure cell without any pressure media. A single crystal used for the pressure measurements, including ambient or zero pressure data, is different than the one used for Figs.~\ref{MRT_all} and \ref{MRT}, hence the ambient pressure data shown here is not the same as shown in these two figures.}
	\label{RTP_warming}
\end{figure}

Figure~\ref{RTP_warming}(a), measured while warming, demonstrates the evolution of the in-plane resistance under pressure, revealing two distinct and noticeable features (similar plots of data taken upon cooling are shown in Fig.~\ref{RTP_cooling} in Section~\ref{PDRJ} of Appendix and demonstrate similar behavior). First, as expected, the transition temperature $T_\mathrm{s}$ increases with increasing pressure, reaching room temperature by $2.2$~GPa. Second, in the intermediate applied pressure of $0< P <1.27$~GPa, the otherwise sharp step-like upturn broadens and contains multiple medium and small sized jumps. The zoomed-in figures in the upper inset, in Figs.~\ref{RTP_warming}(a) and \ref{RTP_cooling}(a), highlight the observed broadening. Such broadening in resistance data is not uncommon in systems with anisotropic changes in the lattice parameters and transition temperature occurring below the solidification temperature of the pressure-transmitting media (PTM). In this scenario, the broadening is due to the non-hydrostatic conditions arising in the system. Broadened features in the resistance measurements under pressure due to non-hydrostaticity were also observed in LaSb$_2$, Ref.~\citenum{Budko_2023}, where the pressure suppresses the charge density wave (CDW) transition. In this study, measurements were conducted using the same pressure cell and PTM as used for \gpt. An extreme example of such solidified pressure medium induced broadening was observed in CaFe$_2$As$_2$ both in orthorhombic-antiferromagentic and collapsed-tetragonal transitions when they occur below the PTM solidification line\cite{Torikachvili_2008,Canfield_2009}.

Figure~\ref{RTP_warming}(b), shown for warming, quantifies the broadening observed in \gpt\ in pressure range of $0< P < 1.2$~GPa; the onset and offset (completion) temperatures of the transition are shown at each applied pressure up to $2.2$~GPa along with the solidification line (SL) of the pressure medium (dotted orange with vertical dash)\cite{Torikachvili_2015}. The P-T phase diagram can be divided into two regions separated by the PTM's SL: region with $T_\mathrm{onset/offset}$ above the solidification line and the one below the line. For transition temperatures above the SL (as well as for the ambient pressure data taken outside the pressure cell) the resistive transition feature is near vertical and the onset and offset temperatures are essentially identical, $\delta T = T_\mathrm{offset} - T_\mathrm{onset} \sim 0$, indicating a sharp and well-defined transition. For transition temperatures below the SL there is a clear difference ($\delta T > 0$) between onset and offset temperatures, thereby confirming and quantifying the broadening. The observed features indicate that the sharp (broadened) features in these measurements are dependent upon whether the transition temperature is above (below) the SL of the PTM, rather than the specific pressure values. Moreover, in \gpt, unlike LaSb$_2$, an increase in pressure beyond $1.2$~GPa, where the transition temperature is higher than the solidification temperature of PTM, leads to a sharpening of the transition feature akin to the measurement conducted at ambient pressure without the use of pressure media. In summary, the broadening arises due to the non-hydrostatic state in the sample resulting from the PTM solidification and anisotropic changes in the lattice parameters, and irrespective of the pressure values, it occurs exclusively when the transition temperature is lower than the solidification temperature of PTM.

In addition, the data in Fig.~\ref{RTP_warming}(b) allow us to quantify the sensitivity of \gpt\ to hydrostatic pressure and make a comparative study with LaSb$_2$\cite{Budko_2023}, where the resistive features were found to be extremely sensitive to pressure with overall pressure derivative of $\sim -500$~K/GPa. The moderately large slope $\dfrac{dT}{dP} \sim 100$~K/GPa for \gpt, obtained from the blue and red data sets, indicates remarkable sensitivity to pressure and implies that the room-temperature crystal structure of \gpt\ is quite fragile.

Additional notable findings are that, unlike in LaSb$_2$, there is almost no change in the width of thermal hysteresis except in the ill-defined broad non-hydrostatic regions. Similarly, in the hydrostatic region there was no significant change in the resistance jump $\dfrac{\delta \rho}{\rho}$, as shown in Fig~\ref{DHystRHOvsP} below in Appendix in the Section~\ref{PDRJ}. Finally, small variation in $\dfrac{\delta \rho}{\rho}$ with pressure in \gpt\ implies that unlike LaSb$_2$, pressure drives the transition temperature to higher values in \gpt\ without bringing any additional significant changes in the Fermi surface or electronic structure.

\section{Discussion\label{DS}}
Temperature dependent measurements of physical properties of three 3A element \xpt\ (X = Al, Ga, and In) single crystals at ambient pressure demonstrated that only \gpt\ exhibits an anomaly corresponding to a structural transition near $80$~K. At this temperature, it undergoes a first-order structural transition from high-temperature $P4/mmm$ to possibly a different crystal symmetry. Based upon the observation of the satellite peaks at two distinct and non-equivalent wave-vector positions, we categorized the low temperature structure as a double-Q type with respect to $P4/mmm$. The observed satellite peaks at (0, 0, 0.5) and (0.5, 0.5, 0.5) suggests doubling of the unit cell along the $c$-direction and in the $\mathbf{ab}$ plane, $a^{\prime} = b^{\prime} = \sqrt{2}a$, below the transition. In addition, \gpt\ shows remarkable sensitivity to pressure implying that the room-temperature structure is fragile: a small perturbation is enough to drive the system towards the structural change. 

Among the members of the 1-5-1 family of compounds, the temperature-driven structural transition has been observed or reported exclusively in \gpt. Given the similar structures reported for all 3A element (X = Al, Ga, In, Tl) \xpt\ members\cite{Boragay_1970,zakharova_2018}, one might reasonably expects similar physical behavior among all. However, our ambient pressure transport and magnetization measurements for \alpt\ and \inpt, shown in Fig.~\ref{MRT_all}, display metallic behavior and diamagnetism, respectively, without any features (anomalies) indicating the structural transition. From the viewpoint of the chemical pressure, one of our consideration was a possibility of the room-temperature structure being different than $P4/mmm$ in these compounds, especially in \alpt\ with the smaller radius atom Al than Ga. However, refinements of our powder X-ray diffraction data, discussed in Section~\ref{RTCS}, confirm the structure of all 3 compounds to be $P4/mmm$.

Furthermore, no additional satellite peaks at $\tau = $ (0, 0, 0.5) and (0.5, 0.5, 0.5) were observed on single crystal X-ray diffraction measurements done on \alpt. Similarly, only satellite peaks at (0, 0, 0.5) positions were measured for \inpt\ and they were absent as well. This confirms that \gpt\ is the sole member of \xpt\ family where the structure transition occurs. Finally, it remains a challenge for computational studies to identify how and why the \gpt\ compound is unique in this manner.

Considering the apparent uniqueness of \gpt, other outstanding questions that remain are as follows: 1) what is the low-temperature crystal symmetry, and 2) what drives the transition? Our DFT calculations provide some insights.

\begin{figure*}[]
	\centering
	\hspace*{-3mm}
	\includegraphics[scale=0.55]{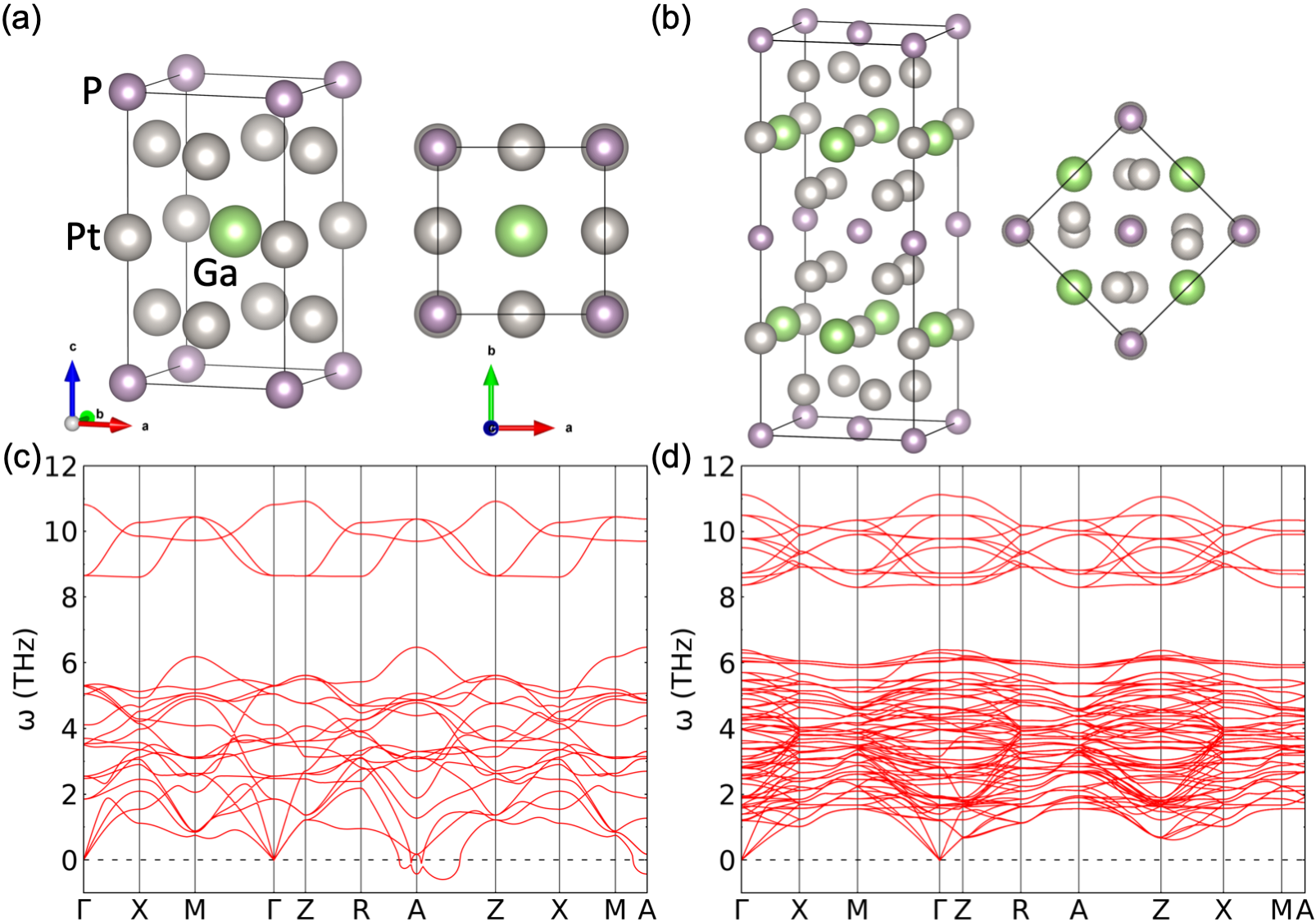}
	\caption{Crystal structures and phonon band dispersions of \gpt\ from the $0$~K DFT calculations. Side-view and top-view along the c-axis of (a) high-T primitive tetragonal structure in space group P4/mmm (123) and (b) low-energy structure of ($\sqrt{2} \times \sqrt{2} \times 2$)R45 supercell. Phonon band dispersion of (c) high-T primitive tetragonal structure and (d) low-energy structure of ($\sqrt{2} \times \sqrt{2} \times 2$)R45 supercell.}
	\label{DFT_phonon}
\end{figure*}

To investigate the structural transition of \gpt\, we carried out phonon calculations at $0$~K with finite displacement method in DFT. As shown in Figs.~\ref{DFT_phonon} (a) and (c), the phonon band dispersion of the \gpt\ primitive tetragonal structure in space group $P4/mmm$ (123) has an imaginary mode at the $A$ point and around it along the $R-A$, $A-Z$ and $M-A$ directions, which means the primitive tetragonal structure is dynamically unstable, agreeing with the experimental observation of the structural change on cooling. By following the eigenvector of the imaginary mode at the $A$ point to double the unit cell vectors and after full relaxation, we find the supercell of ($\sqrt{2} \times \sqrt{2} \times 2$) rotated by $45^\degree$ (R45) is energetically more stable by $30.5$~meV/f.u. Interestingly, as shown in Fig.~\ref{DFT_phonon}(b), the low-energy supercell structure seems to retain all the symmetries, hence should belong to tetragonal space group. Furthermore, in the structure, two out of every four Pt layers has a twist around the c-axis. The phonon band dispersion of the low-energy supercell structure in Fig.~\ref{DFT_phonon}(d) now shows dynamical stability with no imaginary phonon modes, hence, the low-energy supercell structure of ($\sqrt{2} \times \sqrt{2} \times 2$)R45 with lattice parameters $a^{\prime} = b^{\prime} = \sqrt{2}a$ and $c^{\prime} = 2c$ is likley the low-temperature structure. Nevertheless, the experimental determination and confirmation of the crystal structure symmetry at low temperatures are still pending.

Similar phonon calculations done on other two members \alpt\ and \inpt\ lack imaginary modes in their phonon band dispersion, as shown in Fig.~\ref{DPA} in Appendix~\ref{PCAI}, below. The results agree with the experimental observation of no structural changes in these compounds and re-emphasize the association of the imaginary modes with the structural transition in \gpt. 

To summarize, our DFT calculations suggest a potential low-temperature structure and identify softening of phonons as a driving mechanism for the transition. However, this likely captures only a partial aspect of the observed phenomena in \gpt. First, these calculations only account for the presence of superlattice peaks at wave-vector positions (0.5, 0.5, 0.5), leaving the origin of the (0, 0, 0.5) superlattice peaks unexplained. Also, the soft modes are generally invoked for a second order phase transition\cite{Cowley_1980,Krumhans_1992}, unlike the first order transition observed in \gpt. As a result, the origin of these (0, 0, 0.5) superlattice peaks and the underlying reason for the system's double-Q type structural transition remain unanswered questions.

Further experimental and computational investigations are necessary to resolve the ambiguity, completely understand the microscopic mechanisms behind the transition, and identify the uniqueness of \gpt\ that leads to the structural transition. The first step in that direction is to unambiguously determine the low-temperature crystal structure and the refinement of the low-temperature single-crystal X-ray diffraction data would be the appropriate method to achieve this goal. Also, the inelastic X-ray or neutron scattering measurements of phonons with temperature can help to confirm and rule out the softening of phonons at (0.5, 0.5, 0.5) and (0, 0, 0.5) wavevector positions, respectively. 

\section{Summary}

In summary, characterization of single crystals of \xpt\ (X = Al, Ga, In) using various techniques, such as transport, magnetization, and single crystal and powder X-ray diffraction measurements, revealed that \gpt\ exhibits anomalous behavior at low-temperature ($\sim 80$~K) corresponding to a structural transition. No such features were present in \alpt\ and \inpt\ in the temperature range of $2-300$~K. Furthermore, the structural transition in \gpt\ is first order, and is characterized by anisotropic changes in the lattice parameters and volume collapse. Among the members of the 1-5-1 family, \gpt\ stands alone in exhibiting the structural transition. The underlying reason for the unique behavior in \gpt\ remains to be understood, but our DFT calculations offer some insights by suggesting that phonon softening could potentially be one of the mechanisms for the transition. 

Additionally, the pressure dependent measurements of the resistive features indicate that \gpt\ is remarkably sensitive to pressure, and the room-temperature tetragonal $P4/mmm$ structure is fragile. Also, there were noticeable experimental artifacts, such as smeared and broadened transitions, which can be attributed to the non-hydrostatic conditions occurring below the solidification line of the PTM, and irrespective of the pressure values, they occur exclusively when the transition temperature is lower than the solidification temperature of PTM.

\section{Acknowledgement}
Work at Ames National Laboratory was supported by the U.\,S.\ Department of Energy (DOE), Basic Energy Sciences, Division of Materials Sciences \& Engineering, under Contract No.\ DE-AC$02$-$07$CH$11358$. Use of the Advanced Photon Source at Argonne National Laboratory was supported by the U. S. Department of Energy, Office of Science, Office of Basic Energy Sciences, under Contract No. DE-AC02-06CH11357.

\bibliography{GaPtP}

\clearpage

\section{Appendix}

\subsection{Density Functional Theory Calculations\label{DFTC}}

\begin{figure}[h!]
	\centering
	\hspace*{-3mm}
	\includegraphics[scale=0.73]{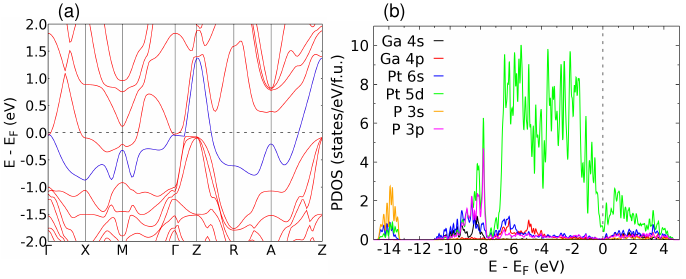}
	\caption{(a) Band structures and (b) Partial density of states for \gpt\ from DFT calculations, corresponding to high-temperature tetragonal $P4/mmm$ structure. The blue line in (a) represents the top valence band according to simple band filling to distinguish the hole and electron pockets.}
	\label{DFTf}
\end{figure}

Electronic structures of \gpt\ from the DFT calculations are shown in Fig.~\ref{DFTf} and features consistent with the 3D metallic behavior are apparent. In (a), band dispersion crossing the Fermi level along all directions except for X-M and R-A are clearly observed and in (b) small but non-negligible Pt $5d$ density of states at $E_\mathrm{F}$ exists. Small density of states near Fermi-level is consistent with the observed diamagnetism.\\

\subsection{Superlattice peaks in different Brillouin zone\label{SPBZ}}
\begin{figure}[H]
	\centering
	\includegraphics[scale=0.75]{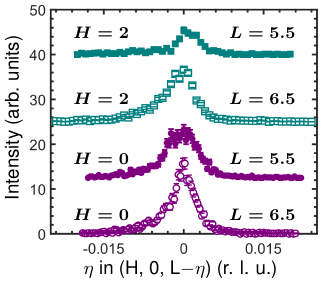}
	\caption{[0, 0, L] scans of the half-integer L peaks at $T = 13$~K for H = 0 and 2.}
	\label{SL_diffBZ}
\end{figure}

Figure~\ref{SL_diffBZ} shows $\tau = (0, 0, 0.5)$ type superlattice peaks in different Brillouin zones.

\subsection{Additional pressure dependence of the resistance features\label{PDRJ}}

\begin{figure}[]
	\centering
	\hspace*{0mm}
	\includegraphics[scale=0.7]{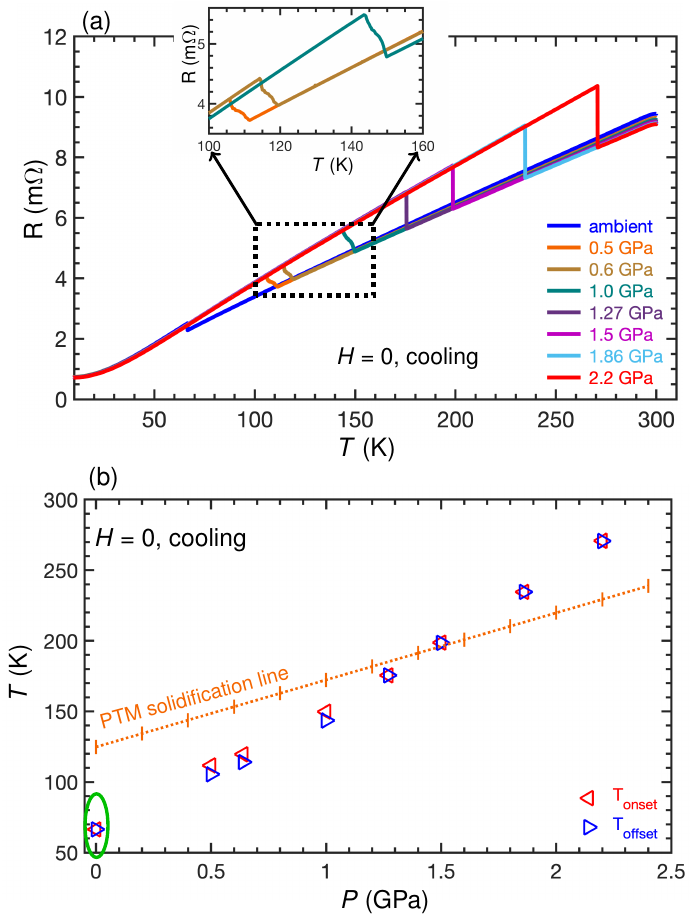}
	\caption{(a) Temperature dependence measurements of the resistance at different pressures illustrating the increase in the feature (upturn or jump) corresponding to the structural transition. Upper inset is a magnified plot, zoomed in to highlight the broadening of an otherwise sharp upturn. (b) Pressure-temperature (P-T) phase diagram of \gpt\ from $H = 0$ resistance measurements showing the onset and offset temperatures in (a). In addition, it contains solidification line (dotted orange with vertical dashes) of the pressure-transmitting media (light oil and n-pentane) at different pressures\cite{Torikachvili_2015}. The green ellipse around the $P \sim 0$~GPa data is to emphasize that the ambient pressure measurements were done outside the pressure cell without any pressure media. A single crystal used for the measurements is different than the one used for Figs.~\ref{MRT_all} and \ref{MRT}.}
	\label{RTP_cooling}
\end{figure}

Figure~\ref{RTP_cooling}(a) demonstrates the evolution of the in-plane resistance under pressure data, which were measured while cooling, and similar to the warming (Fig.~\ref{RTP_warming}(a) in the main text), it demonstrates two distinct and noticeable features, 1) enhancement of the transition temperature $T_\mathrm{s}$ with increasing pressure, 2) multiple medium and small sized jumps for the pressure $0< P <1.27$~GPa. P-T phase diagram in Fig.~\ref{RTP_cooling}(b) reconfirms and quantifies the broadening ($\delta T > 0$) for these pressure values and illustrates that both the onset and offset transitions occur below the solidification lines of PTM.

\begin{figure}[]
	\centering
	\includegraphics[scale=0.73]{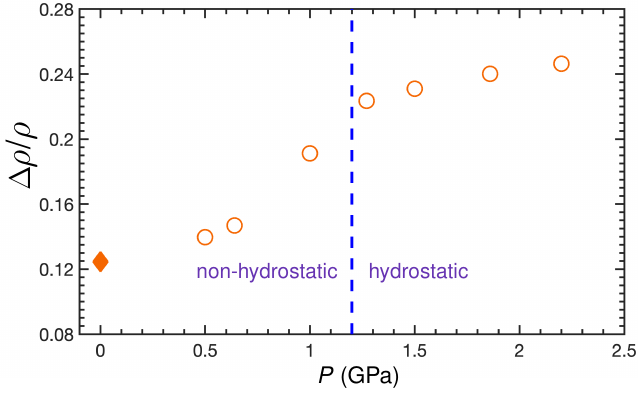}
	\caption{Relative size, $\Delta \rho/\rho$, of the feature in resistance (resistivity) as a function of pressure. The blue dashed line indicates the pressure below which the PTM solidifies and non-hydrostatic conditions are met. The filled diamonds correspond to the ambient pressure measurements outside the pressure cell.}
	\label{DHystRHOvsP}
\end{figure}

$\Delta \rho/\rho$ in Fig.~\ref{DHystRHOvsP} was obtained by methodology similar to that used in Ref.~\citenum{Budko_2023}. Its value nearly doubles (to $24\%$ increase) by $2.2$~GPa but change is very small in comparison to LaSb$_2$, where it increases to $1$ ($100\%$ increase) on increasing pressure to $\sim 0.2$~GPa. Moreover, in LaSb$_2$ the value decreases abruptly in the non-hydrostatic region and eventually disappears. However, no such behavior occurs in \gpt\ except for the slight deviations with a small decrease below $1.27$~GPa in the non-hydrostatic region. Nevertheless, the small and decreased value of $\Delta \rho/\rho$ for the ambient pressure measurement in the figure prevents from making a clear correlation between the decrease and non-hydrostaticity.

Also, the $100\%$ increase in LaSb$_2$ in the hydrostatic region was associated with the further gapping of the Fermi surface\cite{Budko_2023}. Similarly, only small variation in $\Delta \rho/\rho$ in \gpt\ upto $2.2$~GPa implies that pressure on \gpt\ brought no additional significant changes in the Fermi surface or electronic structure.

\subsection{History and ramp-rate dependence of the hysteresis width and transition temperature\label{HDHTT}}

\begin{figure}[]
	\centering
    \hspace*{-5mm}
	\includegraphics[scale=0.75]{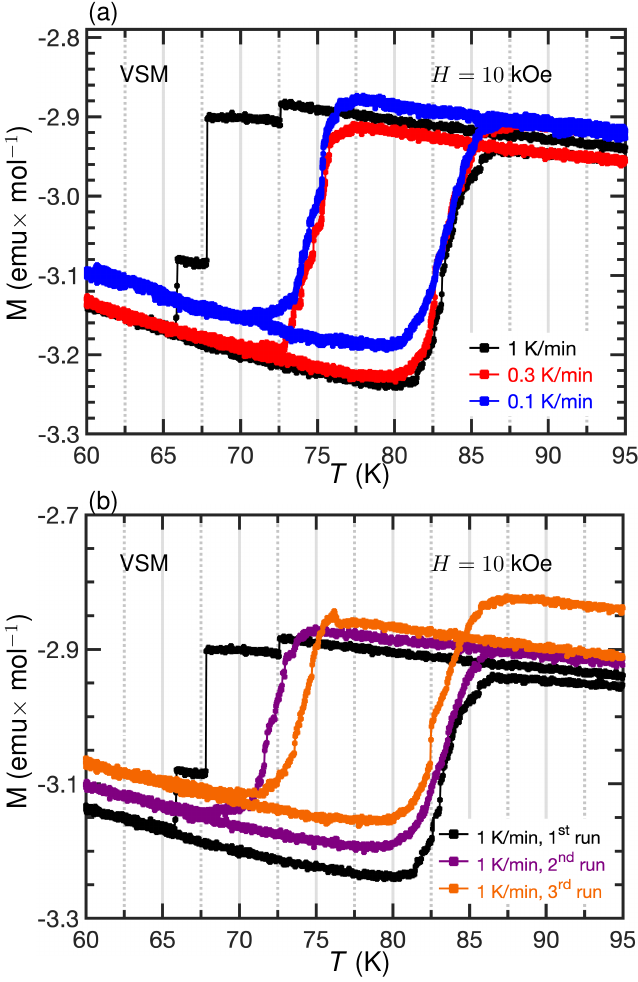}
	\caption{Magnetization measurements done on cluster of \gpt\ single crystals using a VSM option of the MPMS3 with continuous temperature sweeps and field $H = 10$~KOe. Figures illustrate variation in the hysteresis width and transition temperature for measurements done with (a) different ramp-rates and (b) different cycles with the the ramp-rate of 1K/min. In (b) $1^{\mathrm{st}}$ run was the first measurement that started from the room temperature, $2^{\mathrm{nd}}$ run followed it with the measurement up to 180 K, and the $3^{\mathrm{rd}}$ run was measured after leaving the cluster at $180$~K for several hours.}
	\label{MTDC}
\end{figure}

Figure~\ref{MTDC} shows the magnetization measurements done on cluster of \gpt\ single crystals and the measurements were done using a VSM option of the MPMS3 with continuous temperature sweeps. Figure~\ref{MTDC} (a) illustrates that both the hysteresis width and transition temperatures (onset or offset) vary for the measurement done with different cooling(warming) rates. Similarly, Fig.~\ref{MTDC} (b) demonstrates that the variation also occurs for the measurements with the same ramp-rate but with different thermal history. Hence, the first-order transition of \gpt\ exhibits history dependence and this is one of the possible reason why the different measurements discussed above in the main text exhibit slight variations in the hysteresis width and transition temperatures.

\subsection{Phonon band dispersions for \alpt\ and \inpt\ \label{PCAI}}

\begin{figure}[H]
	\centering
	\hspace*{-3mm}
	\includegraphics[scale=1.4]{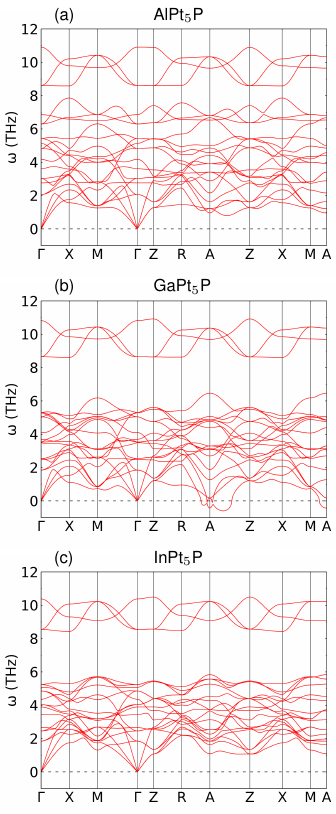}
	\caption{Phonon band dispersion, at 0 K, of high-T primitive tetragonal structure of all 3 members, (a) \alpt, (b) \gpt, and (c) \inpt.}
	\label{DPA}
\end{figure}

Figure~\ref{DPA} shows phonon dispersions, from 0 K DFT calculations, for all 3 members, (a) \alpt, (b) \gpt, and (c) \inpt. The calculations are done for the high-T primitive tetragonal structure of each compounds with their relaxed lattice parameters. The figures clearly show that the imaginary mode in \gpt\ at the $A$ point and around it along the $R-A$, $A-Z$ and $M-A$ directions are absent in both \alpt\ and \inpt. Absence of any imaginary modes in their (\alpt\ and \inpt) phonon dispersion is consistent with the experimental results of no structural transition in these two members. Hence, the imaginary modes related to the lattice instability is the unique feature of \gpt\ among all reported 1-5-1 compounds.

\subsection{Effects of Al and In substitution in GaPt$_5$P \label{AISG}}

The high-pressure resistance measurements show that applying hydrostatic pressure to \gpt\ rapidly enhances the temperature at which the structural phase transition occurs. With this in mind, we were interested to also explore the effect of chemical substitution by alloying \gpt\ with the smaller Al atoms, i.e. \galpt, hoping that this would induce the positive chemical pressure, thereby allowing us to push the transition to higher temperature for more convenient study at ambient pressures.  Likewise, we also attempt to do the opposite and study the effect of negative chemical pressure by alloying with the larger In atoms, \ginpt.

\begin{table}[]
	\caption{In fraction in \ginpt\ estimated from the EDS measurements.\label{InEDS}}
	\begin{ruledtabular} 
		\begin{tabular}{*{6}{|c}|}
            \multirow{2}{*}{Nominal $x$} &
            \multicolumn{5}{c|}{EDS $x$ }\\\cline{2-6}
            & {Spot 1} & {Spot 2} & {Spot 3} & {Spot 4}& Average\\\hline                    
           0.025& 0.0724& 0.0451 & 0.0323 & 0.0295 & 0.04(2)\\
           0.05& 0.0597& 0.0886 & 0.08027 & 0.09162 & 0.08(1)\\
		\end{tabular}
	\end{ruledtabular}
\end{table}

Single crystals of Al and In substituted \gpt\ were grown in the same manner as the pure \gpt\ discussed in the main text but with starting nominal compositions of (Ga$_{1-x}$A$_x$)$_9$Pt$_{71}$P$_{20}$ for A = Al and In and $x = 0.025$ and $0.05$. The heating, cooling, and decanting conditions were otherwise identical to the protocol outlined in the crystal growth section in the main text. EDS measurements were carried out to determine the Al and In levels in the final crystals, and the pure \alpt, \gpt, and \inpt\ were used as standards.

\begin{figure}[]
	\centering
    \hspace*{-15mm}
	\includegraphics[scale=0.72]{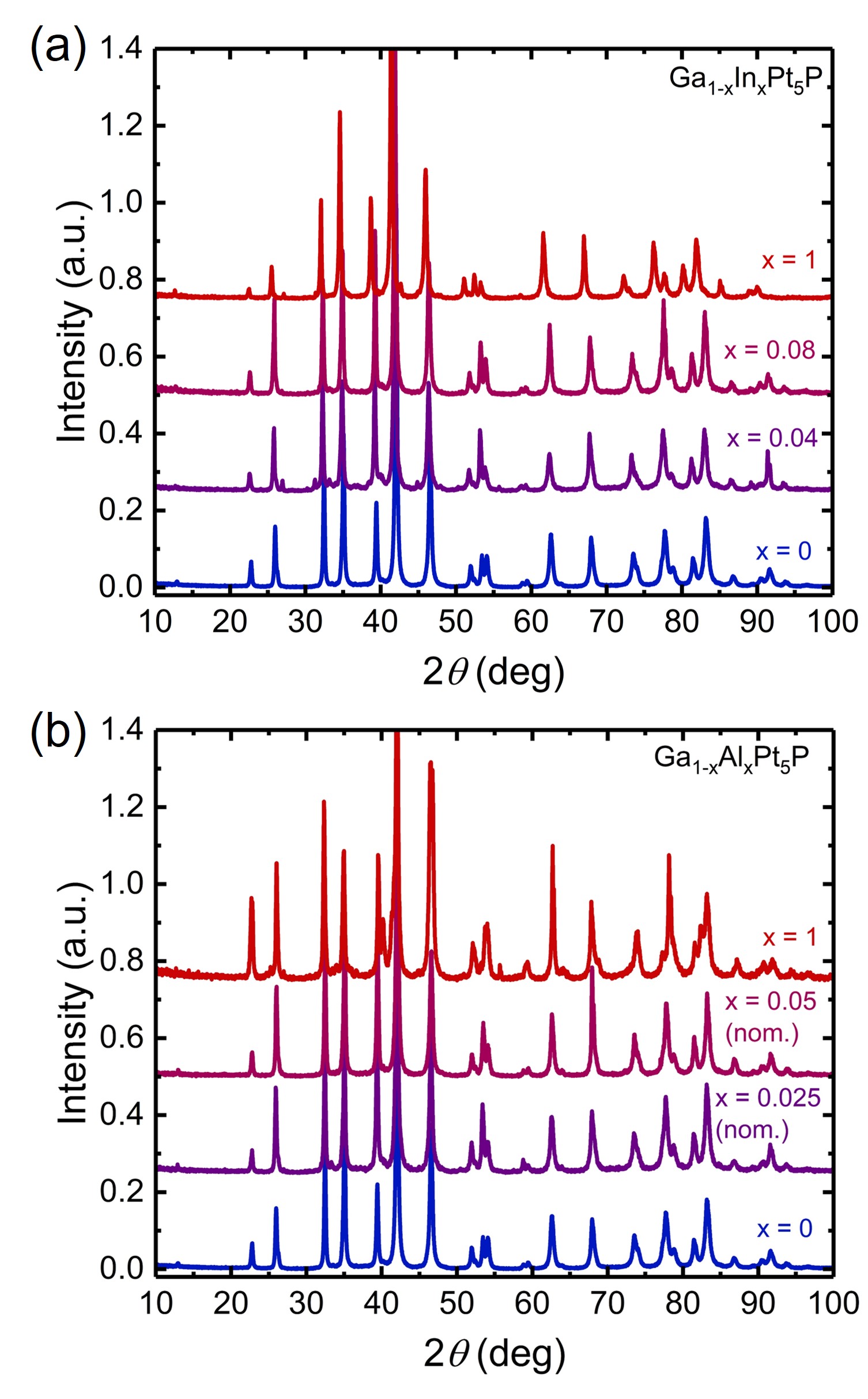}
	\caption{(a) and (b) Powder X-ray diffraction patterns for \ginpt\ and \galpt, respectively.}
	\label{PXRD_Al_In}
\end{figure}

Table~\ref{InEDS} shows the results of EDS measurements on the In alloyed \ginpt. To assess the sample homogeneity, four measurements were made on different spots on the sample surfaces, and we use the average values of each as the measured $x$ in the following discussion. The uncertainties are the standard deviations of the four $x$ values measured for each sample. The EDS measurements show that the measured fraction of In is slightly higher than the nominal values, but that an increasing quantity of In is substituted for Ga as the nominal $x$ increases. In the case of the Al-alloyed samples, we were unable to quantify the alloy fraction of Al, as the measured Al quantity was smaller than the uncertainty in our measurements for both nominally $x = 0.025$ and $x = 0.05$ samples. Therefore, for \galpt, we only refer to the nominal $x$ values in the following discussion to distinguish between samples.

\begin{figure}[]
	\centering
    \hspace*{-5mm}
	\includegraphics[scale=0.43]{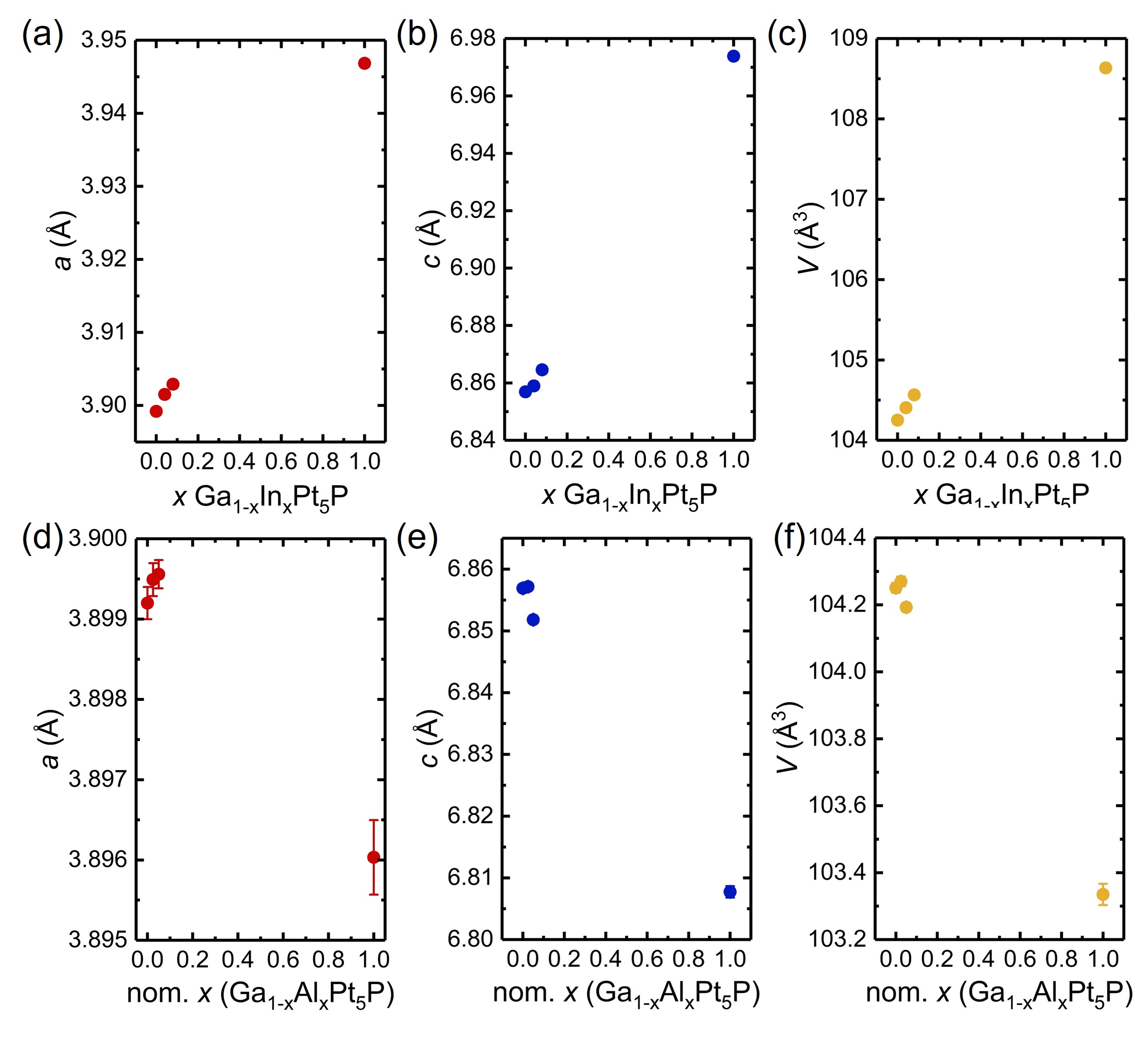}
	\caption{Dependence of the $a$ and $c$ lattice parameters and volume with In composition in \ginpt\ (a--c) and Al composition in \galpt\ (d--f). $X$ corresponds to EDS results for \ginpt\ and nominal concentration for \galpt.}
	\label{LP_Al_In}
\end{figure}

\begin{figure}[]
	\centering
    \hspace*{-5mm}
	\includegraphics[scale=0.6]{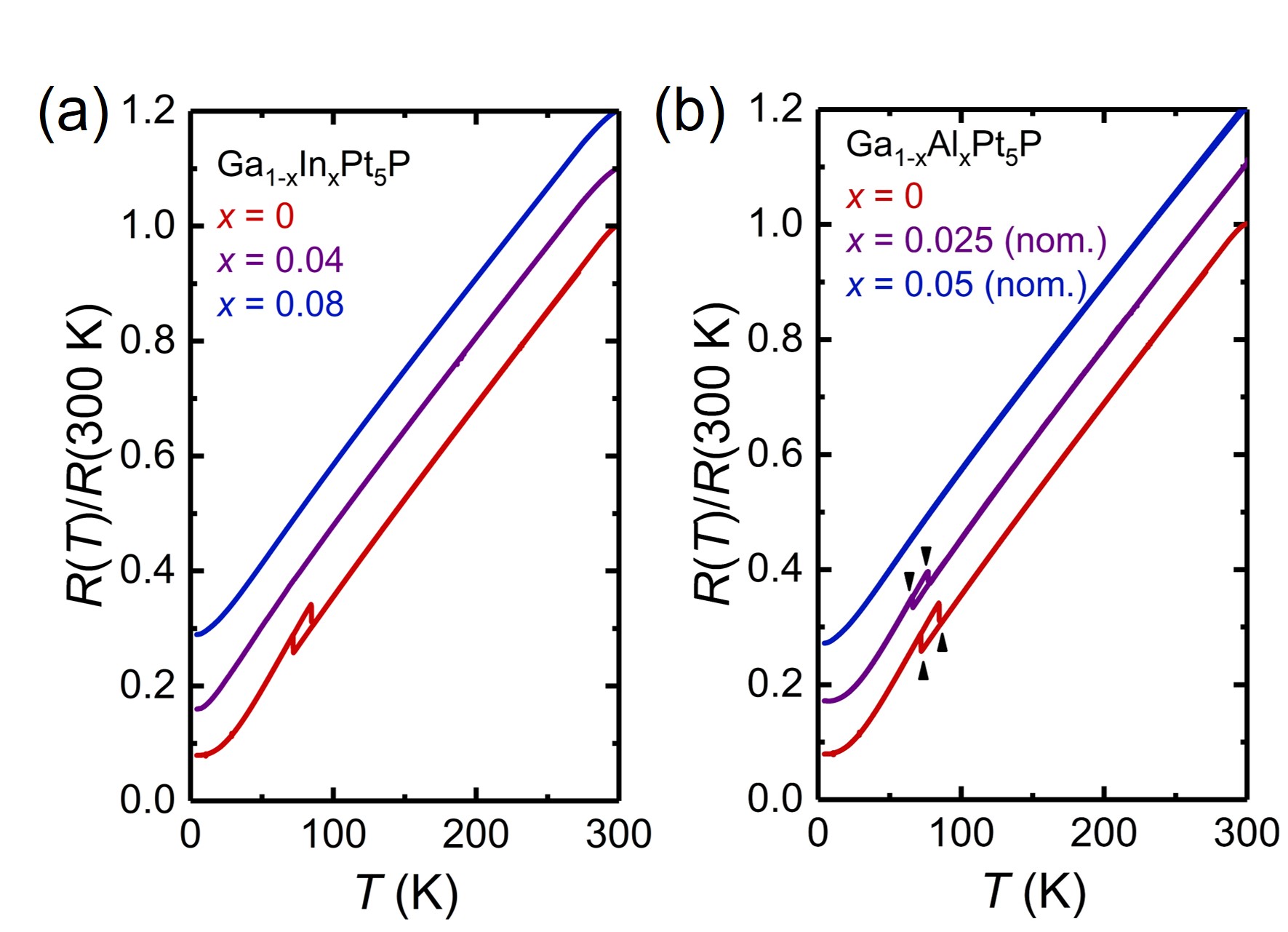}
	\caption{Temperature dependence of R(T)/R(300 K) for 3 different In (a) and Al (b) concentrations, illustrating the rapid suppression of the transition feature with substitution. The Al levels indicated in the figures are the nominal concentration.}
	\label{RvsT_Al_In}
\end{figure}

Figure~\ref{RvsT_Al_In} shows the powder X-ray diffraction patterns for In and Al alloyed \gpt. The measurements were done using our in-house X-ray unit, Rigaku Miniflex-II instrument operating with Cu-$K_\alpha$ radiation, $\lambda = 1.5406$ \AA ($K_{\alpha1}$) and $1.5443$ \AA ($K_{\alpha2}$), at $30$~kV and $15$~mA. In each pattern, the most intense reflections are in good agreement with the expected peaks for the $P$4/$mmm$ structure of \gpt. Several patterns also show a few much weaker reflections that can be indexed to PtP$_2$. The lattice constants refined from the above powder patterns are shown in Figs.~\ref{LP_Al_In} (a)-(c) for In alloyed samples and in (d)-(f) for Al alloyed samples. The lattice parameters of the \ginpt\ samples behave as expected, as $a$, $c$, and the unit cell volume V all increase in a nearly linear manner as the measured In fraction is raised, consistent with the larger radius of In compared to Ga. This result is consistent with the EDS data and suggests a steady increase in In as the nominal $x$ is raised.

In the case of \galpt, we can only compare the measured lattice parameters with the nominal fraction of Al added to the initial growth compositions. We find that for \galpt, the $a$ lattice parameter does not show a uniform trend with nominal doping and increase slightly at low $x$, but shrinks to $\sim 3.896$\AA\ for \alpt. While the initial increase in $a$ at low nominal $x$ appears anomalous, the overall decrease of $a$ from pure \gpt\ to \alpt\ is only $\sim 0.01\%$, meaning that the $a$ lattice parameter is effectively insensitive to Al substitution. On the other hand, $c$ and unit cell volume both decreases by about $1\%$ moving form \gpt\ to \alpt. The changes in the lattice parameters and volume from \gpt\ to \alpt\ are similar to the synchrotron measurement results, listed in Table~\ref{ACALL}.

To determine the evolution of the structural transition in the alloyed samples, we measured the temperature dependence of the resistance, and the results are shown in Figs.~\ref{RvsT_Al_In} (a) and (b) for In and Al alloyed samples, respectively. In both cases, the structural transition is suppressed quickly upon substitution of Ga. The transition is completely absent in both $x = 0.04$ and $x =0.08$ \ginpt\ samples. In the Al doped samples, we find the transition is subtly suppressed from $72$~K (on cooling) or $84$~K (on warming) in pure \gpt\ to $66$~K (on cooling) or $77$~K (on warming) for nominally $x = 0.025$, and vanishes by nominally $x = 0.05$. We note here that despite being unable to resolve the Al fraction with our EDS measurements, the fact that the structure change is clearly suppressed and eliminated for nominally $x = 0.025$ and $x = 0.05$ \galpt\ indicates a possibility of small but finite quantity of Al replacing Ga.

Considering the high-pressure data, these results are surprising. Because application of hydrostatic pressure rapidly increases the transition temperature, we anticipated the smaller Al atoms to provide “positive chemical pressure” and also raise the transition temperature, and the larger In atoms to do the opposite. The data obtained for the In alloyed samples superficially agrees with this prediction. However, given that $4\%$ In leads to a $\sim0.17\%$ volume increase and eliminates the transition, the rate at which the transition vanishes is considerably faster than expected based on “negative chemical pressure” alone. In fact, it is at least twice as fast when considering $\sim0.33\%$ volume change below the transition, shown in Fig.~\ref{acvT}(c). On the other hand, Fig.~\ref{LP_Al_In} (c) shows that the overall unit cell volume of \gpt\ is effectively unchanged by small Al substitution (less than $0.1\%$ at nominally $x = 0.05$), but the structural transition is rapidly suppressed and eliminated in \galpt. Taken together, the data in Fig.~\ref{RvsT_Al_In} indicate 1) that the structural transition in \gpt\ is extraordinarily fragile and can be eliminated with even a relatively gentle perturbation of dilute chemical substitution, and 2) chemical substitution likely has a different effect than applied pressure, and “chemical pressure” is not the correct framework with which to understand the effect of substitution. One possibility is that \gpt\ is very sensitive to disorder, but we do not at present understand the nature of the extreme sensitivity of the transition. Ultimately, more detailed doping and alloying studies beyond the scope of this work are needed to understand the response of \gpt\ to external tuning parameters. 

\end{document}